\documentclass[preprint,eqsecnum,aps,keywords]{revtex4}
\usepackage{dcolumn}
\usepackage{graphicx}
\usepackage{amsmath}
\usepackage{amssymb}
\usepackage{epsfig}
\usepackage{float}
\usepackage{latexsym}
\usepackage{rotating}
\begin{document}
\title{Geodesic flows in  rotating black hole backgrounds}
\author{
Anirvan Dasgupta\footnote{Electronic address: {\em anir@mech.iitkgp.ernet.in}}${}^{}$}
\affiliation{Department of Mechanical Engineering and Centre for Theoretical Studies  \\
Indian Institute of Technology, Kharagpur 721 302, India}
\author{
Hemwati Nandan\footnote{Electronic address: {\em hntheory@yahoo.co.in}}${}^{}$}
\affiliation{Department of Physics  \\
Gurukula Kangri Vishwavidyalaya, Haridwar 249 404, Uttarakhand, India}
\author{
Sayan Kar\footnote{Electronic address: {\em sayan@cts.iitkgp.ernet.in}}${}^{}$}
\affiliation{Department of Physics and Centre for Theoretical Studies  \\
Indian Institute of Technology, Kharagpur 721 302, India}

\begin{abstract}
We study the kinematics of timelike geodesic congruences, in
the spacetime geometry of  rotating black holes in three (the BTZ) and four (the Kerr) dimensions. 
The evolution (Raychaudhuri) equations for the expansion, shear and rotation 
along geodesic flows in the such spacetimes are obtained. For the BTZ case, the equations are solved 
analytically. The effect of the negative cosmological constant on the
evolution of the expansion ($\theta$), for congruences with and without an 
initial rotation ($\omega_0$) is noted. 
Subsequently, the evolution equations, in the case of a Kerr black hole 
in four dimensions are written and solved numerically, for some specific geodesics flows. 
It turns out that, for the Kerr black hole, there exists a critical value of the initial 
expansion below (above) which we have focusing (defocusing). We delineate
the dependencies of the expansion, on the black hole
angular momentum parameter, $a$, as well as on $\omega_0$. 
Further, the role of $a$ and $\omega_0$ on the time (affine parameter) of approach to singularity 
(defocusing/focusing) is studied.
While the role of $\omega_0$ on the time to singularity is as expected, the effect of $a$ leads to 
an interesting new result.
\end{abstract}
\maketitle

\section{Introduction}
Geodesic congruences (flows) and their features in various spacetime
backgrounds have been extensively studied by mathematicians and
physicists alike, over many years. In the context of gravity and general
relativity, studies 
on the evolution (along a chosen geodesic flow) of the
kinematical quantities (the isotropic expansion, shear and rotation (henceforth
referred as ESR)) in specific spacetime geometries, have been carried out.
The ESR evolution, as is well--known, is governed by the Raychaudhuri
equations \cite{wald,toolkit}.
In particular, consequences (geodesic focusing) of the  
Raychaudhuri equation for the expansion scalar are 
crucial while proving the
singularity theorems in general relativity \cite{Penro,Haw}.

Recently, we have initiated a programme where we intend to
study geodesic flows in various contexts (an example outside
the realm of general relativity, is, deformable media)
by solving the evolution equations for the ESR, in various scenarios. 
Some of our work is available in
\cite{adg1}-\cite{adg4}.  

In order to push things further beyond our investigations in \cite{adg3},
here we intend to look at geodesic flows in rotating black hole spacetimes.
To start with, we restrict to three dimensions, where the well-known
black hole solution (with rotation) is the Ba$\tilde{\rm n}$ados--Teitelboim--Zanelli
(BTZ) line element \cite{BTZ,BTZ0}. The BTZ, is a solution of the three dimensional
Einstein equations with a negative cosmological constant. 
Further, in four dimensions, we have the the famous Kerr black
hole \cite{toolkit,wald} which is also important, astrophysically. Such black holes are 
characterised by their mass and spin (rotation rate) and the
study of geodesic flows in their gravitational fields can provide
useful information about the behaviour of matter around strong
gravitational fields. Studies on characterisation of geodesics around a Kerr 
black hole, in terms of exact equatorial homoclinic orbits that separate
bound orbits from plunging, and whirling from non-whirling,
are available in \cite{levin1,levin2}.
The potential usefulness of analysing
families of black hole orbits has been alluded to in \cite{levin2}. 

In this article, we aim towards understanding the kinematics of geodesic 
flows in the above mentioned spacetimes, in three and four dimensions. 
Our article is organised as follows. We first review the BTZ spacetime along 
with its geodesic structure in Section \ref{btzst}. 
The Raychaudhuri equations in the background of the BTZ black hole spacetime 
are 
then written in a freely falling Fermi normal frame. 
Surprisingly, the Raychaudhuri equations in this case are analogous to those 
in the flat background, with stiffness \cite{adg1} identified with the 
modulus of the cosmological constant. The analytical  solutions 
are discussed and the kinematics of flows is revealed accordingly in  
Section \ref{btzrc}.  Next, the Kerr black hole spacetime
geometry (which is a vacuum, axisymmetric solution of the
Einstein equations) with the associated geodesic equations is reviewed 
briefly in Section \ref{4dsbhst} followed by the derivation of 
Raychaudhuri equations in Section  \ref{krayeqn}.
In view of the absence of the analytical solutions of the (geodesic and
Raychaudhuri) equations for the Kerr case, we have solved them numerically.
Based on the evolution of ESR variables for various sets of initial conditions with different values of the other parameters involved, the generic features of 
the geodesic flows are visualised in Section
\ref{krayeqn}. The results are also compared with those obtained for the Schwarzschild
geometry. Finally, in Section \ref{conclu}, we highlight the important results obtained and indicate the possibilities for future work.  

\section{BTZ black hole Spacetime} \label{btzst}

Let us begin by writing down the line element for the BTZ black hole 
in $(2+1)$ dimensions \cite{BTZ,BTZ0},
\begin{equation}
ds^2 = - (N^2-r^2 N_{\phi}^2) \, dt^2 + N^{-2} dr^2  +  r^2 \, d{\phi}^2 + 2 r^2 N_{\phi} dt \, d \phi,
\label{btzmetric}
\end{equation}
Here, the squared lapse function $N^2$ and the angular shift $N_\phi$ are given by,
\begin{equation}
N^2= -M + \frac{r^2}{l^2} + \frac{J^2}{4 r^2}; \,\,\,\,\,\,\,\,\,\,\,\,\,\,\,\, \, \, \, \, \, \, \, \, N_\phi = - \frac{J}{2 r^2}, \label{nnphi}
\end{equation}
with $-\infty <t< \infty, \, 0<r<\infty$, and $0\leq \phi \leq 2 \pi$. 
In equation (\ref{nnphi}), $M$ and $J$ are, respectively, the mass and angular momentum parameters of the BTZ black hole, while  the parameter $l$ is the
radius of curvature of the spacetime geometry, related to the cosmological constant 
($\Lambda$) as $l^2 = -1/\Lambda$.  The lapse function vanishes for the 
following two values of $r$,
\begin{equation}
r_{\pm}= l \left[ \frac{M}{2}\, \left(1 \pm \sqrt{1- \frac{J^2}{M^2 l^2}}\right) \right]^{\frac{1}{2}},
\end{equation}
where $r_{\pm}$ identifies the horizons of the BTZ black hole, 
which exist only when the following conditions are satisfied,
\begin{equation}
M > 0; \, \, \, \, \, \,\,\,\,\,\,\,\,\,\,\,\,\,\,\,\,\,\,\,\,\,\,\,\,\vert J \vert \leq  M l.
\end{equation}
The inner ($r_{-}$) and outer ($r_{+}$)  horizons shrink with the increasing 
value of the cosmological constant. One may also note here that both the 
roots of the lapse function coincide for the extremal case $\vert J \vert = M l$. The first integral of the geodesic equations corresponding to the line element (\ref{btzmetric}) are given as follows \cite{Cruz, Farina},
 \begin{equation}
\dot t =   - \frac{1}{N^2}\, (E +N_\phi L),
\label{tdot}
\end{equation}
\begin{equation}
\dot \phi  = \frac{N_\phi}{N^2} \,(E+ N_\phi L) - \frac{L}{r^2},
\label{phidot}
\end{equation}
\begin{equation}
\dot r= \pm \left[(E+ N_\phi L)^2 -\frac{N^2 L^2}{r^2} - N^2 \right]^{\frac{1}{2}},
\label{rdot}
\end{equation}
where $E$ and $L$ are integration constants which represent the energy and angular momentum, respectively. Here $u^i = (\dot t, \dot r, \dot \phi)$ satisfies 
the timelike constraint $ u^i u_i = -1$ for the timelike geodesics. 
The equations (\ref{tdot})-(\ref{rdot}) characterise the motion of the 
test particles in the BTZ black hole spacetime.

\section{Evolution of ESR in the local frame} \label{btzrc}

\subsection{Fermi normal frame and the Raychaudhuri equations}
We construct an orthonormal Fermi frame with $u^i, {\hat e}^i_\alpha$ ($\alpha=1,2$) where
\begin{equation}
g_{ij} {\hat e}^i_\alpha {\hat e}^j_\beta = \delta_{\alpha\beta} \hspace{0.2in}
;\hspace{0.2in} g_{ij} u^i {\hat e}^j_{\alpha} =0
\end{equation} 
This is the local inertial frame of a freely falling observer. The
basis set $\left \{ u^i, {\hat e}^{i}_\alpha \right \}$ is parallely transported along the geodesic. 


The tensor $B_{ij}=\nabla_j u_i$,  defined in the
coordinate frame, may be rewritten in the local frame of the freely 
falling observer  as $B_{\alpha \beta} = B_{ij} {\hat e}_{\alpha}^{i}   {\hat e}_{\beta}^{j}$. Further, $ B_{\alpha \beta}$ can be decomposed into its 
trace, symmetric traceless and antisymmetric parts in the usual way,
\begin{equation}
B_{\alpha \beta} = \frac{1}{2} \theta \, \delta_{\alpha \beta} + \sigma_{\alpha \beta} + \omega_{\alpha \beta}, \label{bdecom}
\end{equation}
where,
\begin{equation}
\frac{1}{2}\theta \delta_{\alpha \beta}=
\left(
\begin{array}{cc}
\frac{1}{2}\theta & 0\\
0&  \frac{1}{2}\theta\\
\end{array} \right),~~~~~~
\sigma_{\alpha \beta}=
\left(
\begin{array}{cc}
\sigma_{+} & \sigma_{\times} \\
\sigma_{\times} & -\sigma_{+} \\
\end{array} \right), ~~~~~~~
\omega_{\alpha \beta}=
\left(
\begin{array}{cc}
0 & -\omega \\
\omega& 0 \\
\end{array} \right).
 \label{comp}
\end{equation}
In the above, $\theta $ is the expansion scalar, $\sigma_{+}$ and $ \sigma_{\times}$ are the components of trace--free shear tensor $
\sigma _{\alpha \beta}$ while $\omega$ denotes the component of rotation 
in the antisymmetric tensor $\omega_{\alpha \beta}$. The evolution equation for $B_{\alpha \beta}$ in the Fermi normal frame as discussed above is then given as,
\begin{equation}
\dot{B}_{\alpha \beta} + B_{\alpha \gamma} B^{\gamma}_{\,\,\,\,\beta}= - R_{ikjl}\, u^k u ^l \,{\hat e}_\alpha^{i} {\hat e}_\beta^{j}. \label{bevol1}
\end{equation}
Since we are working in three dimensions, we can make use of the
following relation between the Riemann, Ricci tensors and the Ricci scalar,
\begin{equation}
R_{ikjl} = g_{ij} R_{kl} + g_{kl}R_{ij} - g_{kj} R_{il}-g_{il} R_{jk} -
\frac{1}{2} \left ( g_{ij} g_{kl} - g_{il} g_{jk} \right ) R
\end{equation}
and also the relation $R_{ij} = \Lambda g_{ij}$.
Consequently, the equation for $B_{\alpha\beta}$ becomes
\begin{equation}
\dot{B}_{\alpha \beta} + B_{\alpha \gamma} B^{\gamma}_{\beta}= \Lambda \delta_{\alpha\beta}. \label{bevol}
\end{equation}
Using (\ref{bdecom}), and (\ref{bevol}), the Raychaudhuri equations for 
the ESR variables turn out to be,
\begin{equation}
\dot \theta + \frac{1}{2}{\theta^2} + 2 (\sigma_{+}^{2} +\sigma_{\times}^{2} -\omega^2) + 2\, \vert \Lambda \vert =0,
\label{theta}
\end{equation}
\begin{equation}
\dot \sigma_{+} +\theta \sigma_{+} =0,
\label{sigma+}
\end{equation}
\begin{equation}
\dot \sigma_{\times}+ \theta \sigma_{\times} =0,
\label{sigmacross}
\end{equation}
\begin{equation}
\dot \omega+ \theta \omega=0
.\label{omega}
\end{equation}
The above first--order, coupled, nonlinear and inhomogeneous 
ordinary differenial equations 
have to be solved, for given initial conditions, in order to obtain 
the kinematical evolution of the ESR along 
geodesic congruences.  

A subtle point may be noted here. Given the (static) velocity vector field $u^i=(\dot t,\dot \phi,\dot r)$ in (\ref{tdot})-(\ref{rdot})
one can easily find the ESR variables as functions of $r$ using the definition $B_{ij}=\nabla_ju_i$. 
This leads to, for example, 
\begin{equation}
\theta = \nabla_i u^i = \pm \frac{l^2(E^2 +M) -L^2 -2 r^2}{l^2 r\left[ - \frac{4 L^2}{l^2}+ (E^2 +M) + 
\frac{4 L^2 M - J(4 LE +J)}{r^2} -\frac{4r^2}{l^2}\right]^{1/2}}, \label{thdirect}
\end{equation}
and the vorticity $\omega\equiv 0$.
Physically, this represents
the kinematics of a geodesic congruence in which all geodesics have 
a fixed value of the constants $E$ and $L$. Therefore, 
this expression cannot accommodate arbitrary initial conditions on the velocity field of a congruence, or
on the ESR variables. The solutions obtained by integrating the Raychaudhuri equations (\ref{theta})-(\ref{omega}) directly, for given arbitrary initial
conditions, are, in 
this sense more general, and thereby, underscores the importance of these equations.

 \subsection{Analytical solutions }

The equations (\ref{theta})-(\ref{omega}) have been 
exactly solved analytically (see \cite{adg1,adg2} for the detailed procedure). It is interesting to note from equations (\ref{theta})-(\ref{omega}) here
 and (2.8)-(2.11) in  \cite{adg1} that, kinematically, the 
equations in the BTZ black hole background can be mapped to those for 
flows in a flat space with isotropic stiffness $\vert \Lambda \vert$ ($k$ in \cite{adg1}).
In order to solve this set of equations,  we first define a parameter 
$I = \sigma_{+}^{2} +\sigma_{\times}^{2} -\omega^2$ and obtain the following 
analytical expressions for the ESR for the three different cases depending on the signature of $I$.\\
 \noindent {\bf Case A}: $I>0$
 \begin{equation}
\theta= \frac{ a p \cos (c_1 + \frac{p t}{2})}{2\,[a \,
\sin (c_1 + \frac{p t}{2}) + b]},
\label{thetaigreaterzero}
\end{equation}
\begin{equation}
\{\sigma_{+},\sigma_{\times},\omega \}= \frac{\{c_2,c_3,c_4\}}{[a \, \sin (c_1 + \frac{p
t}{2}) + b]},
\label{common}
\end{equation}
 where $ p^2 = 16 \vert \Lambda \vert $ and $ a^2 = 16 +b^2$. Here we have used $\{ \, \}$ brackets to express the solutions in a compact way for the components of shear and rotation which have a common denominator as in equation (\ref{common}). The parameter $b$ and the integration constants ($c_1,c_2,c_3$ and $c_4$) involved can be given in terms of the initial values of ESR (i.e., $\theta_0,\sigma_{+0},\,\sigma_{\times 0}$ and $\omega_{0}$) as described below,\\
\begin{equation}
b = \sqrt {\frac{I_0}{p^2}}
\left( -8 + \frac{2\theta_0^{2}}{I_0} +\frac{p^2}{2I_0}\right ), \label{beqn}
\end{equation}
\begin{equation}
c_1 = \tan^{-1} \left(\frac{p-b\sqrt I_0}{2\theta_0}\right), 
\end{equation}
\begin{equation}
\{ c_2,\, c_3,\, c_4\} = \{\,\sigma_{+0},\,\sigma_{\times 0},\, \omega_{0}\, \}\, (a \, \sin c_1 + b).
\end{equation}

\noindent {\bf Case B}: $I=0$\\
\begin{equation}
\theta = 2 \sqrt{\vert \Lambda \vert } \tan [\sqrt{\vert \Lambda \vert} (c_1-t)], \end{equation}
\begin{equation}
\{\sigma_{+},\sigma_{\times},\omega \}= \{c_2,c_3,c_4\} \sec^2 [\sqrt{\vert \Lambda \vert} (c_1-t)].\end{equation}
The integration constants in terms of the initial ESR for this case are as given below,
\begin{equation}
c_1 = \frac{1}{\sqrt {\vert \Lambda \vert}} \, \tan^{-1} \left (\frac{\theta_0}{2\sqrt {  \vert \Lambda \vert}} \right ),
\end{equation}
\begin{equation}
\{\,c_2,\, c_3,\, c_4\,\} = \frac{\{\,\sigma_{+0},\,\sigma_{\times 0},\,  \omega_{0}\,\}}{\sec^2 (\sqrt{ \vert \Lambda \vert}\, c_1)} .
\end{equation}

 {\bf Case C}: $I<0$\\
\begin{equation}
\theta= \frac{ a p \sin (c_1 + \frac{p t}{2})}{2\,[a \,
\cos (c_1 + \frac{p t}{2}) + b]},
\label{thetailesszero}
\end{equation}
\begin{equation}
\{\sigma_{+},\sigma_{\times},\omega \}= \frac{\{c_2,c_3,c_4\}}{[ a \, \cos (c_1 + \frac{p
t}{2}) + b]}.
\label{common1}
\end{equation}
Here $b$ is the same as in equation (\ref{beqn}) for the case $I>0$ and the 
integration constants in terms of initial values of the ESR are,
\begin{equation}
c_1 = \tan^{-1} \left(\frac{2\theta_0}{p -b\sqrt{ - I_0}}\right), 
\end{equation}
\begin{equation}
\{\, c_2,\,  c_3,\, c_4 \,\} = \{\, \sigma_{+0},\, \sigma_{\times 0}, \, \omega_{0}\, \} \, (a \, \cos c_1 + b).
\end{equation}
\begin{figure}[h]
\includegraphics*[height=5.5cm,width=12.5cm]{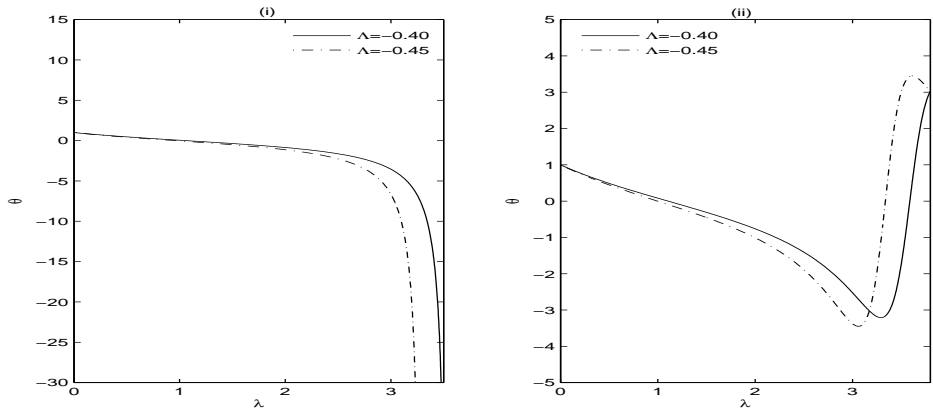}
\caption{Effect of $\Lambda$ on the evolution of the expansion scalar $\theta(\lambda)$ for the cases (i) 
without initial rotation, and (ii) with initial rotation ($\omega_0=0.2$).}
\label{fig1btz}
\end{figure}
\begin{figure}[h]
\includegraphics*[height=5.5cm,width=11.5cm]{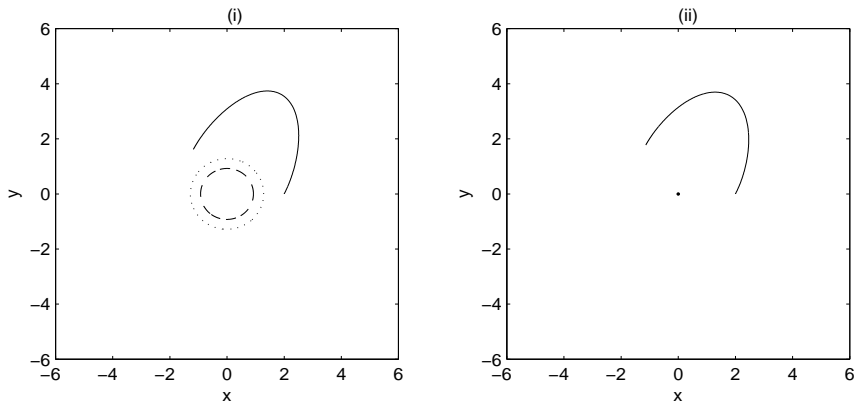}
\caption{Orbits around the BTZ black hole without initial rotation for (i) with horizon ($\Lambda=-0.4$), and (ii) without horizon ($\Lambda=-0.45$). 
The dotted and dashed lines indicate, respectively, the outer 
$(r_{+})$ and inner $(r_{-})$ horizons in the left panel above. }
\label{fig2btz}
\end{figure}
\begin{figure}[h]
\includegraphics*[height=5.5cm,width=11.5cm]{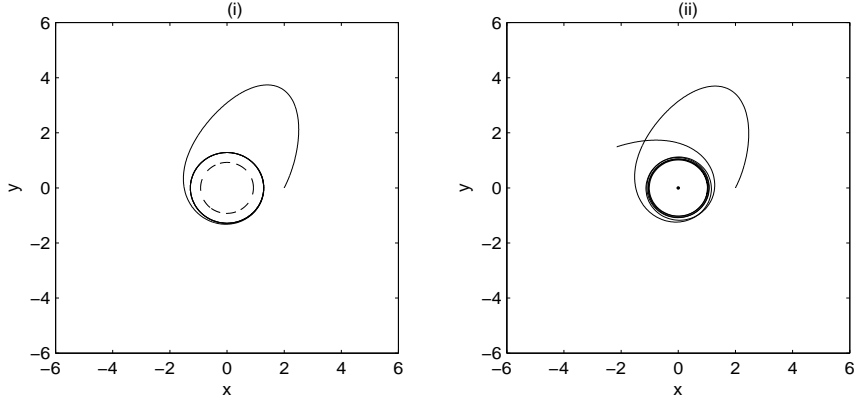}
\caption{Orbits around the BTZ black hole  with initial rotation ($\omega_0=0.2$) for (i) with horizon ($\Lambda=-0.4$), and (ii) without horizon ($\Lambda=-0.45$).}
\label{fig3btz}
\end{figure}
All the  solutions mentioned above become singular in finite time. Note, however, that
this is a {\em singularity of the congruence}.  
One may note that the kinematics of geodesic flows, as 
governed by the Raychaudhuri equations, in the freely falling frame, 
is independent of the value of the 
parameters $M$, $J$, $E$ and $L$. 
However, the geodesics are dependent on these parameters and we use the 
following values of parameters $M=1, E=2, L=2$ and $J=1.5$ 
for visualising the orbits around the BTZ black hole. 
For the kinematics, we consider two cases 
with $\Lambda = -0.40$  and $\Lambda =-0.45$ which correspond to 
the BTZ black hole with and without 
horizon, respectively. For these two cases, we observe the effect of rotation 
on the expansion scalar 
$\theta$ in Fig.\ref{fig1btz}. The corresponding orbits are presented in  Fig.\ref{fig2btz} (without initial rotation) and 
Fig.\ref{fig3btz} (with initial rotation). It is observed that the geodesic congruence is necessarily singular without initial 
rotation (see Fig.\ref{fig1btz}), a fact which can also be inferred from 
the analytical solutions. 
Geodesic focusing always occurs earlier with the shrinking of the horizon of the
BTZ black hole, due to an increase in the value of $|\Lambda|$, 
as shown in Fig.\ref{fig1btz}(i). However beyond a
critical value of the initial rotation, the congruence 
singularity is removed irrespective of the presence/absence of the horizon, as shown in Fig.\ref{fig1btz}(ii). 
It may be noted that the Fig.\ref{fig1btz}(i) corresponds to the exact solutions of the ESR variables represented 
by equations (\ref{thetaigreaterzero}) and (\ref{common}) for the case $I>0$. Further, Fig.\ref{fig1btz}(ii) 
corresponds to the exact solutions of the ESR variables represented by equations (\ref{thetailesszero}) and (\ref{common1}) for the case $I<0$. 

It might seem that the expression for $\theta$ in (\ref{thdirect})
does have a dependence on the parameters $M$, $J$ $E$, $L$ among others. However, one can show using the solutions of the
geodesic equations (i.e., $r(\lambda)$) in (\ref{thdirect}) that the final result coincides with
one of the solutions obtained above in this section. Thus, it
is surprising to note that, while the geodesic trajectories depend on the parameters $M$, $J$, etc.,
the ESR variables depend only on $|\Lambda|$. This is essentially an
effect of the spacetime dimension ($2+1$ here) where the Weyl tensor
is identically zero and no explicit dependence on $M$, $J$ appear in the
ESR evolution equations.

\section{The Kerr black hole spacetime} \label{4dsbhst}
In this section, we quickly recall the well-known Kerr
black hole spacetime (in the Boyer--Lindquist coordinates, 
which are analogous to the Schwarzschild coordinates 
for a non--rotating black hole) with mass $M$ and
angular momentum $J$ as characterised by the following stationary and
axisymmetric metric,
\begin{eqnarray}
ds^2 = -\left( 1-\frac{2 M \, r}{\rho^2}\right)\,  dt^2 \,  - \, \frac{4 M \, a \, r
\sin^2\psi}{\rho^2} \, d \phi \, dt \, + \,
\frac{\rho^2}{\Delta} \, dr^2 + \rho^2 \,d\psi^2 \nonumber
\\
+ \left(\,r^2 + a^2 + \frac{2 \,M \,r \,a^2 \,\sin^2
\psi}{\rho^2}\,\right) \, \sin^2\psi \, d \phi^2,
\label{kerrmetric}
\end{eqnarray}
Here $ a=J/M$, $\rho^2= r^2+a^2 cos^2\psi$ and
$\Delta=r^2-2Mr+a^2$. With $a=0$, the Kerr metric
(\ref{kerrmetric}) reduces to the Schwarzschild. 
The geodesic equations corresponding to the
metric  (\ref{kerrmetric}) are well--known  (we do not repeat them here). 
However, analytical expressions of first integrals of the geodesic equations 
are not known for the general case (except for the choice of $\psi=\pi/2$ \cite{Hartle}). 
In our work here, we are interested in timelike geodesic congruences 
for which the constraint $u^\alpha u_\alpha=-1$ is satisfied, i.e.,
\begin{equation}
g_{00} \dot t^2+g_{11} \dot r^2+g_{22} \dot \psi^2 +g_{33} \dot \phi^2 + 2g_{03}\dot t \dot \phi +1=0. \label{tlc}
\end{equation}
Once the initial conditions on the velocity vector components are specified 
in such a way that this constraint is satisfied, it will be
satisfied throughout the geodesic evolution.
\section{ESR evolution} \label{krayeqn}
\subsection{The Raychaudhuri equations}
Consider a congruence of geodesics with the 
associated time like vector field $u^\alpha$, tangent to the geodesic at 
each point and satisfying equation (\ref{tlc}). 
A transverse metric $h_{\alpha \beta}$ on a spacelike hypersurface 
can be written by decomposing the spacetime metric $(g_{\alpha \beta})$ 
as defined through equation (\ref{kerrmetric}) into 
the longitudinal ($-u_\alpha u_\beta$) and transverse part 
(i.e., $h_{\alpha \beta})$,
\begin{equation}
h_{\alpha \beta}=  g_{\alpha \beta}+ u_{\alpha} u_{\beta};\, \, \, \, \, \, \,
\, \, \, \, \, \, \, \, \, \, \, \, (\alpha, \beta = 0,1,2,3).\label{gab}
\end{equation}
The transverse metric satisfies $u^\alpha h_{\alpha \beta} = 0 $ which means 
that  $h_{\alpha \beta}$ is orthogonal to $u^{\alpha}$ and the 
transverse spacelike hypersurface represents the local rest frame of a 
freely falling observer in the given spacetime. 
The evolution of the ESR on such a transverse hypersurface, 
can be investigated by using a tensor $B_{\alpha \beta}$, which can 
be decomposed into its trace, symmetric traceless and antisymmetric parts 
as follows,
\begin{equation}
B_{\alpha \beta} = \frac{1}{3}{\theta} \, h_{\alpha \beta} +\sigma_{\alpha \beta} + \omega_{\alpha \beta}, \label{bij3d}
\end{equation}
where $\theta = B^{\alpha}_{\, \, \alpha}$ is the expansion scalar,
$\sigma_{\alpha \beta} = B_{(\alpha \beta)} - \theta \, h_{\alpha \beta} /3 $
the shear tensor, and $ \omega_{\alpha \beta} = B_{[\alpha \beta ]}$, the
rotation tensor with the brackets $( \, )$ and $[\, ]$ denoting  
symmetrisation and antisymmetrisation, respectively.  The shear and rotation 
tensors also satisfy  $ h^{\alpha \beta}\, \sigma_{\alpha \beta} = 0 $ 
and $ h^{ \alpha \beta}\, \omega_{\alpha \beta} = 0$. 
The evolution equation for $B_{\alpha \beta}$ takes the following form,
\begin{equation}
\dot{B}_{\alpha \beta} + B_{\alpha \gamma} B^\gamma_{\,\,\,\beta} =
- R_{\alpha \eta \beta \delta} \, \, u^{\eta}u^{\delta},
\label{dotb1}
\end{equation}
Using the equation (\ref{dotb1}) and following standard methods 
\cite{toolkit,adg3,review}, the Raychaudhuri equations for the ESR variables
in the Kerr black hole spacetime, given by equation (\ref{kerrmetric}) 
turn out to be,
\begin{equation}
\dot \theta + \frac{1}{3}{\theta^2} + \sigma^2 -\omega^2
= 0, \label{theta3e}
\end{equation}
\begin{equation}
\dot \sigma_{\alpha \beta} + \frac{2}{3} \theta \, \sigma_ {\alpha
\beta} +  \sigma_ {\alpha \gamma} \sigma^{\gamma}_{\, \, \beta} +
\omega_ {\alpha \gamma} \omega^{\gamma}_{\, \, \beta} +
\frac{1}{3} (\sigma^2 -\omega^2) \, h_{\alpha \beta} + C_{\alpha
\eta \beta \delta}\, u^{\eta} \, u^{\delta} = 0, \label{sigma3}
\end{equation}
\begin{equation}
\dot \omega _{\alpha \beta} +  \frac{2}{3} \theta \,
\omega_{\alpha \beta} + \sigma_{\alpha}^{\, \, \gamma} \,
\omega_{\gamma \beta} + \omega_{\alpha}^{\, \, \gamma} \, \sigma_{
\gamma \beta} = 0, \label{omega3}
\end{equation}
where $\sigma^2 = \sigma_ {\alpha \beta} \,\sigma^{\alpha \beta}$,  
$\omega^2 = \omega_ {\alpha \beta} \,\omega^{\alpha \beta}$ and  
$C_{\alpha \eta \beta \delta} \equiv R_{\alpha
\eta \beta \delta}$ is the Weyl tensor for the present case.  
One can notice the absence of the geometric part
 (i.e., $R_{\alpha \beta} \, u^{\alpha} \,
u^{\beta}$ ) in equation (\ref{theta3e}) since the Kerr spacetime
is Ricci flat. Further, the nonzero Weyl in equation (5.5), with its
dependence on the rotation parameter $a$ (and also $M$) will lead to
a dependence on $a$ in the evolution of the ESR.   

\subsection{Visualisation of ESR evolution}
\noindent  
In the absence of analytical solutions for the ESR variables in this case, we have directly solved (\ref{dotb1}) numerically.
In order to represent these kinematical quantities at any point, we consider a freely falling (Fermi) normal frame with the 
basis vectors $E_{\mu}^\alpha$, $\mu=0,\ldots,3$  
(with $E_{0}^\alpha = \hat{u}^{\alpha}$) which are parallely transported. To construct
such frames numerically, we solve the differential equations $u^\beta\nabla_\beta \, E_{\mu}^{\alpha} =0$ 
(with initial conditions of an orthonormal frame) simultaneously with (\ref{dotb1}) \cite{adg3}.  
The tensor $B_{\alpha \beta}$ in the Fermi basis may be represented as,
$$
B_{\alpha \beta} = (\frac{1}{3}\theta +\sigma_{11}) e^1_{\alpha}e^1_{\beta} + (\frac{1}{3}\theta +\sigma_{22}) e^2_{\alpha}e^2_{\beta} + 
(\frac{1}{3}\theta -\sigma_{11}-\sigma_{22}) e^3_{\alpha}e^3_{\beta} + (\sigma_{12} - \omega_3)e^1_{\alpha} e^2_{\beta} +  \nonumber $$
\begin{equation}
(\sigma_{21} + \omega_3)e^2_{\alpha}e^1_{\beta} + (\sigma_{13} + \omega_2)e^1_{\alpha}e^{3}_{\beta} + (\sigma_{31} - 
\omega_2) e^{3}_{\alpha}e^1_{\beta}+ (\sigma_{23} - \omega_1) e^{2}_{\alpha}e^{3}_{\beta}+ (\sigma_{32} + \omega_1) e^{3}_{\alpha}e^{2}_{\beta}  \, ,\label{bij4d}
\end{equation}
where $e^\mu_{\alpha}$ are co-frame basis which satisfy the relation  
$e^\mu_{\alpha} E_\nu^{\alpha} = \delta^{\mu}_{\nu}$. The ESR variables 
can now be constructed from the evolution tensor (\ref{bij4d}), using the basis vectors, as,
\begin{equation}
\theta = B_{\alpha \beta}\, h^{\alpha \beta} \equiv B_{\alpha \beta} g^{\alpha \beta}, \label{theta1}
\end{equation}
\begin{equation}
\sigma_{11}= B_{\alpha \beta}\,  E_{1}^{\alpha}\, E_{1}^{\beta} -\frac{1}{3} \theta,
\label{sig11}
\end{equation}
\begin{equation}
\sigma_{22}=  B_{\alpha \beta}\,  E_{2}^{\alpha}\, E_{2}^{\beta} -\frac{1}{3} \theta,
\label{sig22}
\end{equation}
\begin{equation}
\sigma_{12} = \frac{1}{2} (B_{\alpha \beta} \, E_{1}^{\alpha}\, E_{2}^{\beta} + B_{\alpha \beta}\,  E_{2}^{\alpha}\, E_{1}^{\beta}),
\label{sig12}
\end{equation}
\begin{equation}
\sigma_{13}=\frac{1}{2} ( B_{\alpha \beta} \, E_{1}^{\alpha}\, E_{3}^{\beta} + B_{\alpha \beta}\,  E_{3}^{\alpha}\, E_{1}^{\beta}),
\label{sig13}
\end{equation}
\begin{equation}
\sigma_{23}= \frac{1}{2} (B_{\alpha \beta} \, E_{2}^{\alpha}\, E_{3}^{\beta} + B_{\alpha \beta}\,  E_{3}^{\alpha}\, E_{2}^{\beta}),
\label{sig23}
\end{equation}
\begin{equation}
\omega_1= \frac{1}{2} (B_{\alpha \beta} \, E_{3}^{\alpha}\, E_{2}^{\beta} - B_{\alpha \beta}\,  E_{2}^{\alpha}\, E_{3}^{\beta}),
\label{om1}
\end{equation}
\begin{equation}
\omega_2= \frac{1}{2} (B_{\alpha \beta} \, E_{1}^{\alpha}\, E_{3}^{\beta} - B_{\alpha \beta}\,  E_{3}^{\alpha}\, E_{1}^{\beta}),
\label{om2}
\end{equation}
\begin{equation}
\omega_3= \frac{1}{2} (B_{\alpha \beta} \, E_{2}^{\alpha}\, E_{1}^{\beta} - B_{\alpha \beta}\,  E_{1}^{\alpha}\, E_{2}^{\beta}).
\label{om3}
\end{equation}

In order to understand the  focusing/defocusing  behaviour of a 
timelike geodesic congruence, 
let us redefine the expansion scalar as $\theta = 3{\dot F}/F$ where the dot 
indicates the derivative with respect to
the affine parameter $\lambda$. The equation (\ref{theta3e}) can now
be written in the following Hill-type equation,
\begin{equation}
{\ddot F} + H \, F = 0, \label{theta3e1}
\end{equation}
where $H =(\sigma^2-\omega^2)/3$ with $\sigma^2= 2(\sigma_{11}^2 +
\sigma_{22}^2 + \sigma_{12}^2 + \sigma_{13}^2 + \sigma_{23}^2 + \sigma_{11}
\sigma_{22})$ and $\omega^2 = 2(\omega_1^2 +\omega_2^2 +\omega_3^2)$. 
If $F\rightarrow 0$ in finite time, we have a finite time singularity in 
$\theta$ with focusing (defocusing) if 
$\dot{F}<0$  ($\dot{F}>0$).
The signature of the invariant quantity $H$ is useful in understanding
the individual roles of shear and rotation in the occurrence of geodesic focusing/defocusing.
When $H$ is positive definite (i.e., $\sigma^2 >\omega^2$), 
there exists conjugate points and geodesic 
focusing/defocusing takes place. 
On the other hand, no finite time singularity exists for an initially 
non-contracting congruence 
(i.e., $\theta_0\ge 0$) if $H$ is negative definite. For $\theta_0<0$, there exists a critical value
below which focusing/defocusing will take place.

\begin{figure}[h]
\includegraphics*[width=14.5cm]{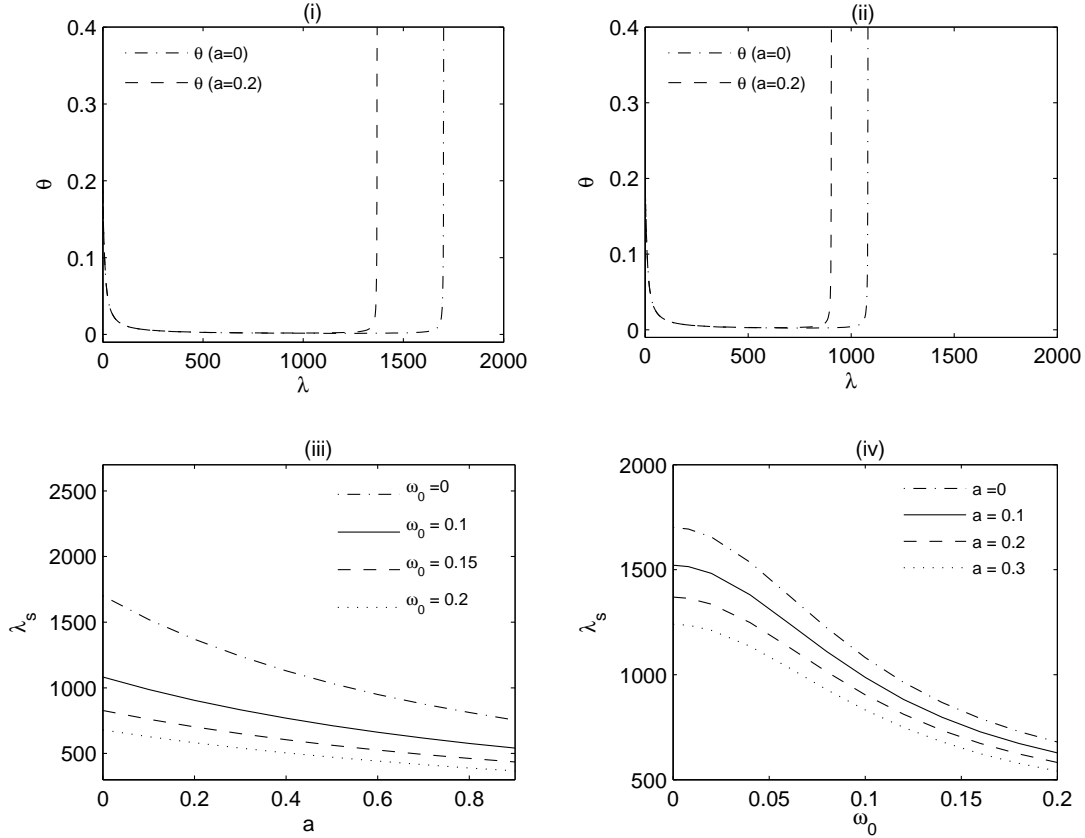}
\caption{Defocusing (for $\theta_0>\theta_c$) for the cases (i) without initial rotation, and (ii) with initial rotation $(\omega_0= 0.1)$. 
The effect of (iii) angular momentum, and (iv) initial rotation on time of approach to singularity ($\lambda_s$).  The initial shear in all cases is zero.}
\label{fig1}
\end{figure}
 \begin{figure}[h]
\includegraphics*[width=14.5cm]{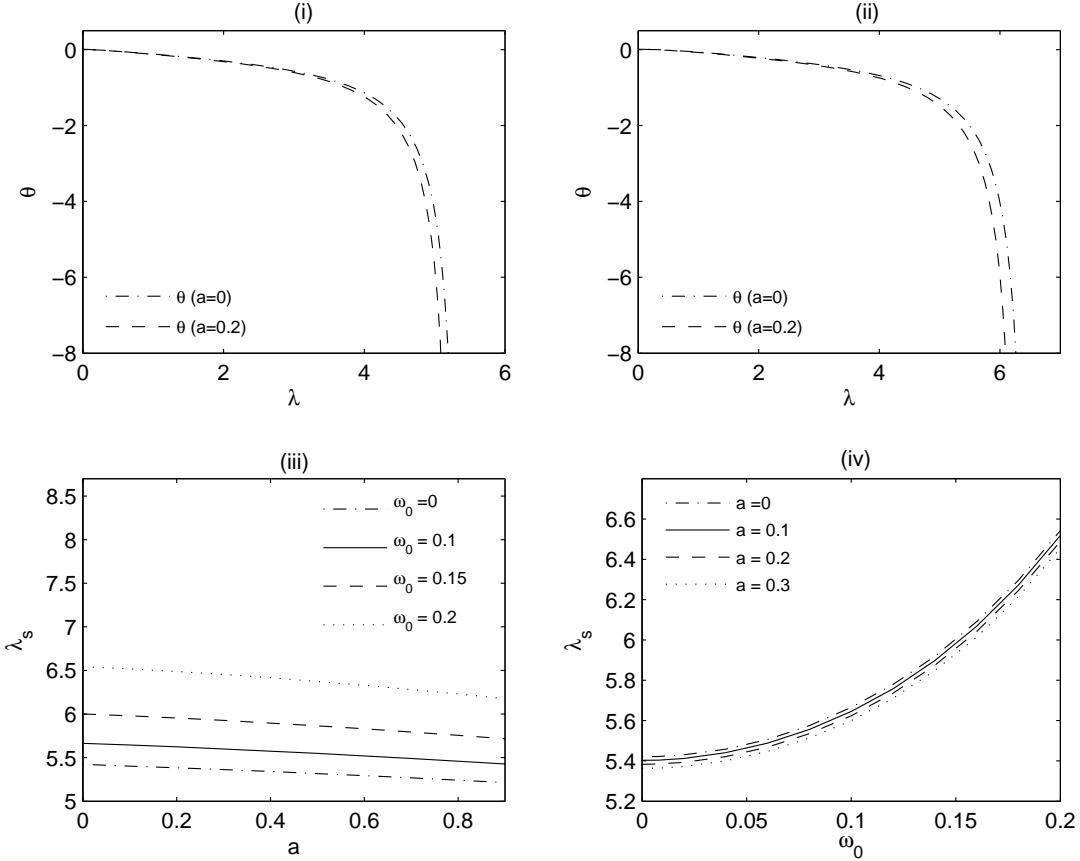}
\caption{Focusing (for $\theta_0<\theta_c$) for the cases (i) without initial rotation, and (ii) with initial rotation $(\omega_0= 0.2)$.  The effect of (iii) angular momentum, and (iv) initial rotation on time of approach to singularity ($\lambda_s$).  The initial shear is $\sigma_0=0.28$.}
\label{fig2}
\end{figure}
In the following, the evolution of the ESR variables in the Kerr black hole
 background, under different conditions, 
are presented and compared with those in the Schwarzschild background for 
some interesting cases.

We consider the case when $\psi=\pi/2$ (equatorial section). 
We have observed that for an initially expanding congruence, geodesic defocusing takes place
 (i.e., $\theta \to \infty)$ and there exists a critical value of the expansion
 scalar $\theta_c$ below which there is a focusing  (i.e., $\theta \to -\infty)$. 
It is important to mention that with
 the change in the initial conditions on shear and rotation, the critical
value of the initial expansion scalar will also change. 
The results with no initial shear and rotation are presented in Figs. \ref{fig1}(i) and (ii) for $\theta_0$ above the critical
initial expansion (defocusing), and Figs. \ref{fig2}(i) and (ii) for $\theta_0$ below the critical initial expansion (focusing)
for both Kerr and Schwarzschild ($a=0$) black holes. 
In order to see the effects of the angular momentum $a$ and the 
initial rotation $\omega_0$ of the congruence, 
we have calculated the time to singularity $\lambda_s$ for the super-critical and sub-critical values of $\theta_0$
and presented them, respectively, in Figs. \ref{fig1}(iii) and (iv), and Figs. \ref{fig2}(iii) and (iv). 
For the case in Fig. \ref{fig1}, we observe that both angular momentum and 
initial rotation of the congruence assist defocusing. On the other hand 
in Fig. \ref{fig2},
it is observed that with increase in $\omega_0$, the congruence focuses at 
a later instant, as expected, while an opposite trend is observed for 
increase in $a$. Thus, we conclude that the angular momentum $a$ advances the time
to singularity (irrespective of its type), while the initial rotation $\omega_0$ of
the congruence advances defocusing and delays focusing.
For this reason, for the same set of initial conditions, focusing/defocusing occurs 
earlier in the Kerr black hole as compared to the Schwarzschild. 
We have also noticed that the effect of initial
shear on geodesic focusing and defocusing for the Kerr case is the 
same as that for the Schwarzschild (results are not presented here). 
Focusing occurs at an earlier instant of time when we have some
initial shear in comparison to the case with zero initial shear.

Since all values of $M$ and $a$ do not correspond to a Kerr black hole
and the horizon exists only when $a \leq M$, the limiting 
value of the angular momentum is $a=M$ (extreme Kerr). 
We have observed that for the extreme Kerr, all the features 
of the geodesic focusing and defocusing remain qualitatively similar as 
those for the non--extreme Kerr case, though with a 
change in the scale.

\section{Summary and Conclusions} \label{conclu}
In this article, we have investigated the kinematics of 
of timelike geodesic congruences in the background of the BTZ and  Kerr black hole spacetimes.
The  important conclusions are summarised below:
\begin{itemize}
\item{Apart from the cosmological constant ($\Lambda$), the other
parameters of the BTZ black hole spacetime, do not have any direct role 
in the kinematics of the geodesic congruence. However, the geodesic trajectories
surely depend on the black hole parameters.}
\item{An increase in $\Lambda$ in the BTZ case, which implies a larger negative curvature, 
enhances focusing of the congruence. } 
\item{The notion of focusing remains qualitatively similar in the BTZ 
black hole background for the cases with or without horizons.}
\item{
The angular momentum paramter $a$ of the Kerr black hole assists both focusing and defocusing when the respective cases occur. 
Thus, physically, the larger the spin of the black hole, the earlier the focusing/defocusing 
of the geodesic congruence occurs. This effect is absent in the BTZ black hole case. } 
\item{In contrast with the role of $a$, 
an increase in the initial rotation, $\omega_0$, of the congruence
assists defocusing but delays focusing in both the $(2+1)$ and $(3+1)$ dimensional rotating black holes. 
This observation is similar to results obtained previously \cite{adg1}-\cite{adg3}.}

\end{itemize}

The tidal distortion due to the presence of mass nearby, 
or accretion of matter into a  rotating black hole deserves careful attention. 
The kinematics of inflows can be conveniently studied using the
Raychaudhuri equations in these backgrounds.
The effect of the black hole angular momentum parameter on the focussing/defocussing of 
geodesic congruences, as has been observed in this
work, can have an interesting role in the accreation process and may produce observable effects.
Further, for a more comprehensive understanding, the kinematics of flows
in maximally extended coordinate systems, for black hole spacetimes, 
would be interesting to investigate.
Finally, the study of kinematics of null geodesic flows in such spacetimes 
is surely worth trying out in future.

\section*{Acknowledgments}
\noindent The authors sincerely thank the Department of Science
and Technology (DST), Government of India for financial support
through  a sponsored project (grant number: SR/S2/HEP-10/2005). 
One of the authors (HN) would also like to acknowledges the financial 
support from University Grants
Commission (UGC), New Delhi, India in
terms of the UGC-Dr. D. S. Kothari Post-Doctoral
Fellowship during part of this work at the Centre of Theoretical Physics, Jamia Millia Islamia, New Delhi. HN is also thankful to the Centre for Theoretical Studies (CTS), Indian Institute of Technology, Kharagpur for warm hospitality under CTS Visitors' Programme.


\begin{references}
\bibitem{toolkit} E. Poisson, {\em A relativists' toolkit: the mathematics
of black hole mechanics}, (Cambridge University Press, 2004).
\bibitem{wald} R. M. Wald, {\em General Relativity}, (University of Chicago
Press, Chicago, USA, 1984).
\bibitem{levin1} J. Levin and G. Perez-Giz, {\em Homoclinic orbits around spinning black holes. I. Exact solution for the Kerr separatrix}, Phys. Rev. {\bf D79} (2009)
124013 (arXiv:0811.3814).
\bibitem{levin2} G. Perez-Giz and J. Levin, {\em Homoclinic orbits around spinning black holes. II. The phase space potrait}, Phys. Rev. {\bf D79} (2009)
124014 (arXiv:0811.3815).
\bibitem{Penro} R. Penrose, {\em Gravitational collapse and space-time singularities}, Phys. Rev. Lett. {\bf 14} (1965) 57.
\bibitem{Haw} S. W. Hawking,{\em Occurrence of singularities in open universes}, Phys. Rev. Lett. {\bf 15} (1965) 689; {\em Singularities in the universe},ibid {\bf 17} (1966) 444 .
\bibitem{adg1} A. Dasgupta, H. Nandan and S. Kar, {\em Kinematics of deformable media}, Annals Phys. {\bf 323} (2008) 1621 (arXiv:0709.0582).
\bibitem{adg2} A. Dasgupta, H. Nandan and S. Kar, {\em Kinematics of flows on curved, deformable media}, Int. J. Geom. Meth. Mod. Phys. {\bf 6} (2009) 645.
 (arXiv:0804.4089 [physics.class-ph]).
\bibitem{adg3} A. Dasgupta, H. Nandan and S. Kar, {\em Kinematics of geodesic flows in stringy black hole backgrounds},  Phys. Rev. {\bf D79} (2009) 124004
(arXiv:0809.3074 [physics.class-ph]).
\bibitem{adg4} S. Ghosh, A. Dasgupta, and S. Kar, {\em Geodesic congruences in warped spacetimes},  Phys. Rev. {\bf D83} (2011) 084001
(arXiv:1008.5008 [gr-qc]).
\bibitem{review} S. Kar and S. SenGupta, Pramana,  {\bf 69} (2007) 49; gr-qc/0611123 and references therein.
\bibitem{BTZ} M. Ba$\tilde{\rm n}$ados, C. Teitelboim and J. Zanelli, {\em The black hole in three dimensional spacetime}, Phys. Rev. Lett. {\bf 69} (1993) 1849 (arXiv:hep-th/9204099v3).
\bibitem{BTZ0} M. Ba$\tilde{\rm n}$ados, C. Teitelboim, C. Henneaux and J. Zanelli, {\em  Geometry of the (2+1) black hole}, Phys. Rev.{\bf D48} (1992) 1506.
 
\bibitem{Cruz} N. Cruz, C. Mart$\acute{\rm i}$nez and L. Pe$\tilde{\rm n}$a, {\em Geodesic structure of the (2+1)-dimensional BTZ black hole},  Class. Quantum Grav. {\bf 11} (1994) 2731. 
\bibitem{Farina} C. Farina, J. Gamboa, Antonio J. Segui-Santonja, {\em Motion and trajectories of particles around three-dimensional black holes}, Class. Quantum Grav. {\bf 10} (1993) L193. 
\bibitem{Hartle} J. M. Hartle, {\em Gravity  An Introduction to Einstein's General Relativity} (Pearson Education Inc., Singapore 2003).
\end{references}
\end{document}